\documentclass[oneside,english]{elsart}
\usepackage[T1]{fontenc}
\usepackage[latin1]{inputenc}
\usepackage{graphicx}

\makeatletter

\usepackage{babel}
\makeatother
\begin{document}
\begin{frontmatter}

\title{On the propensity of magnetism in $3d$ transition-metal-$MgCNi_{3}$
alloys}

\author{P. Jiji Thomas Joseph and Prabhakar P. Singh}

\address{Department of Physics, Indian Institute of Technology  Bombay, Powai,
Mumbai-400076, India}

\begin{abstract}
The changes in the electronic properties of the substitutionally disordered
$MgC(Ni_{1-x}T_{x})_{3}$ ($T$$\equiv$$Fe$, $Co$ or $Cu$) alloys
are studied using the atomic sphere formulation of the Korringa-Kohn-Rostoker
coherent-potential approximation method (KKR-ASA CPA), while the effects
of incipient magnetism in these alloys are studied phenomenologically
using Ginzburg-Landau coefficients in conjunction with fixed-spin
moment method. We find that the disordered $MgC(Ni_{1-x}T_{x})_{3}$
alloys have a small magnetic moment localized at $Fe$ and $Co$ sites
for low concentrations. The overestimation of the calculated magnetic
moment is likely to be due to the limitations of the local-density
approximation used in the present study. However, the calculated Ginzburg-Landau
coefficients clearly show that the disordered $MgC(Ni_{1-x}T_{x})_{3}$
alloys remain paramagnetic. At expanded volumes, we also find the
possibility of a ferromagnetic state for $MgC(Ni_{0.95}Fe_{0.05})_{3}$
and $MgC(Ni_{0.90}Co_{0.10})_{3}$.
\end{abstract}
\end{frontmatter}
Incipient magnetism akin to spin-fluctuations is anticipated \cite{PRB-64-140507,PRL-257601}
in the $8$$K$ superconductor $MgCNi_{3}$ \cite{Nature-411-54}.
A strong van Hove singularity just below the Fermi energy $E_{F}$
\cite{PRL-027001,PRB-172507} in $MgCNi_{3}$ suggests that the material
may be close to a magnetic instability. Using a rigid band picture,
it is estimated that approximately $0.5$ holes would make $MgCNi_{3}$
magnetic \cite{PRL-027001,PRB-100508}. However, in experiments the
hole dopings accomplished via $Fe$ or $Co$ substitutions in the
$Ni$ sub-lattice of $MgCNi_{3}$ do not lead to magnetic solutions
\cite{PRB-064510,SSC-132-379,SSC-491,BJP-32-755,PRB-064503}. 

Model Hamiltonian- based calculations \cite{SSC-122-269} suggest
that dilute impurities of $3d$ transition-metals would develop a
small magnetic moment in $MgCNi_{3}$. Other calculations \cite{PRB-064525,SSC-132-379,PRB-66-024520,PhysicaC-154,JAP-91-8504},
based either on supercells or on  ordered $MgC(Ni_{1-x}T_{x})_{3}$
alloys with $T$$\equiv$$Fe$ or $Co$, infer that in low concentrations
the impurities would remain paramagnetic. The results of the Model
Hamiltonian-based calculations are inconsistent with the experiments,
at least for $Fe$ and $Co$. The results of other calculations using
a rigid-band-like picture, although consistent with the experiments
in so far as the lack of magnetism is concerned, do not provide a
reliable and a complete description of substitutionally disordered
$MgC(Ni_{1-x}T_{x})_{3}$ alloys. The present study is intended to
overcome some of the shortcomings of the previous studies and provide
a reliable and a detailed description of the effects of $3d$ transition-metal
impurities in $MgCNi_{3}$. 

In the present study of the substitutionally disordered $MgC(Ni_{1-x}T_{x})_{3}$
alloys, with $T$$\equiv$$Fe$, $Co$ or $Cu$ and $0\leq x\leq1$,
the effects of chemical disorder are taken into account by employing
the first-principles approach of Korringa-Kohn-Rostoker in the atomic
sphere approximation in conjunction with coherent-potential approximation
(KKR-ASA CPA) \cite{ProgMater-27-1}. The effects of incipient magnetism
in $MgC(Ni_{1-x}T_{x})_{3}$ alloys are studied phenomenologically
using Ginzburg-Landau coefficients obtained from total energy versus
magnetic moment calculations carried out with fixed-spin moment method. 

To obtain the Ginzburg-Landau coefficients one first calculates, using
the fixed-spin moment method, the magnetic energy $\Delta E(M)=E(M)-E(0)$,
where $E(0)$ and $E(M)$ are the total energies for the paramagnetic
phase and ferromagnetic phase with a net magnetic moment $M$ respectively,
for $MgC(Ni_{1-x}T_{x})_{3}$ alloys. Then the calculated $\Delta E(M)$
as a function of $M$ is fitted to the form ${\displaystyle {\displaystyle \sum_{n>0}\frac{1}{2n}a_{2n}M^{2n}}}$
for $n=4$. The fit coefficients $a_{2n}$ can be used to study the
propensity of magnetism in the substitutionally disordered $MgC(Ni_{1-x}T_{x})_{3}$
alloys as a function of $x$. For example, in the phenomenological
approach the coefficient $a_{2}<0$ would represent a ferromagnetic
ground state while $a_{2}>0$ would imply a paramagnetic state.

Before describing our work in detail, we briefly describe the main
results of the present study. We find that the disordered $MgC(Ni_{1-x}T_{x})_{3}$
alloys have a small magnetic moment localized at $Fe$ and $Co$ sites
for low concentrations. The overestimation of the calculated magnetic
moment is likely to be due to the limitations of the local-density
approximation. From the calculated Ginzburg-Landau coefficients we
conclude that the disordered $MgC(Ni_{1-x}T_{x})_{3}$ alloys remain
paramagnetic throughout the concentration range. At expanded volumes,
we also find the possibility of a ferromagnetic state for $MgC(Ni_{0.95}Fe_{0.05})_{3}$
and $MgC(Ni_{0.90}Co_{0.10})_{3}$.

The unpolarized, spin polarized, and fixed-spin moment calculations
for $MgC(Ni_{1-x}T_{x})_{3}$ alloys are carried out, self-consistently,
using the KKR-ASA-CPA method. The calculations are scalar-relativistic
with the partial waves expanded up to $l_{max}=3$ inside the atomic
spheres. The exchange-correlation effects are taken into account using
the Perdew and Wang parametrization \cite{PRB-45-13244}. The core
states have been recalculated after each iteration. To improve the
ASA calculations, we have incorporated both the muffin-tin correction
for the Madelung energy \cite{PRL-55-600,PRB-51-5773} and the multipole
moment correction to the Madelung potential and energy \cite{CMC-15-119}
in our calculations. For this purpose the multipole moments of the
electron density have been determined up to $l_{max}^{M}=6$. Further,
screening constants $\alpha$ and $\beta$ were included in our calculations
following the prescription of Ruban and Skriver \cite{PRB-66-024201,PRB-66-024202}.
These values were estimated from the order(N)- locally self-consistent
Green's function method \cite{PRB-56-9319} and were determined to
be $0.83$ and $1.18$, respectively. The atomic sphere radii of $Mg$,
$C$ and $Ni/Co$ were kept as $1.404$, $0.747$, and $0.849$ of
the Wigner-Seitz radius, respectively. The overlap volume resulting
from the blow up of the muffin-tin spheres was less than $15$\%.
The integration of the Green's function over the energy to evaluate
moments of the density of electronic states was carried out along
a semi circular contour comprising of $20$ points in the complex
plane. For the Brillouin zone integrations, the special $\mathbf{k}-$point
technique was employed with $1771$ $\mathbf{k}-$points spread in
the irreducible wedge of the cubic Brillouin zone. 

For $MgCNi_{3}$ the equilibrium state is found to be paramagnetic,
in agreement with the previous reports \cite{PRB-64-140507,PRB-100508,PRB-172507,PRL-027001}.
The calculated equilibrium lattice constant and bulk modulus for $MgCNi_{3}$
were found to be $7.139\,$$a.u.$ and $0.42\, Mbar$ respectively,
in agreement with earlier theoretical calculations \cite{PRB-64-140507,PRB-100508}.
The density of states at Fermi energy $E_{F}$ for $MgCNi_{3}$ evaluated
at the calculated equilibrium lattice constant is $14.56\, state/Ry-atom$
$(5.35\, state/eV-cell)$. This value has been found to be sensitive
to the approximations involved in the band structure calculations,
and have been reported to be in the range of $4.8$-$6.4$ $state/eV-cell$
\cite{PRB-64-140507,PRB-100508,PRB-172507,PRL-027001}. 

For $MgC(Ni_{1-x}T_{x})_{3}$ alloys for $0.0\leq x\leq0.2$, the
lattice constants decreased linearly for both $Fe$ and $Co$ substitutions
at the rates of $-0.076$ and $-0.073$ $a.u./at$\%, respectively,
while for $Cu$ substitutions the lattice constants increased at the
rate of$+0.164$$a.u./at$\%. 

\begin{figure}[h]
\includegraphics[%
  clip,
  scale=0.3,
  angle=-90]{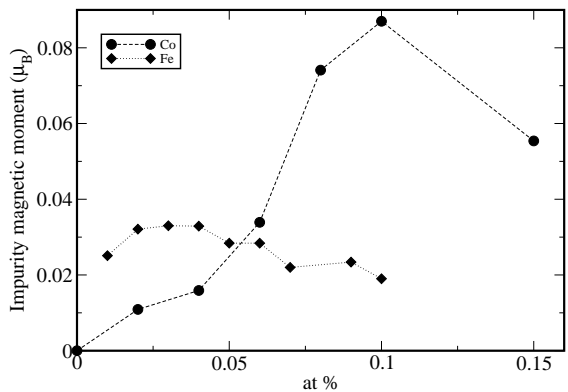}

\caption{\label{LocMagMom}The local magnetic moments at the $Fe$ site in
$MgC(Ni_{1-x}Fe_{x})_{3}$ and $Co$ site in $MgC(Ni_{1-x}Co_{x})_{3}$
as a function of $x$, calculated as described in the text. }
\end{figure}

The spin polarized calculations for disordered $MgC(Ni_{1-x}T_{x})_{3}$
alloys showed a small but finite local magnetic moments at both $Fe$
and $Co$ sites in their corresponding disordered alloys. Note that
the total energies of the unpolarized calculations are essentially
degenerate with the corresponding total energies of the spin polarized
calculations. The variation of the local magnetic moment in $MgC(Ni_{1-x}T_{x})_{3}$
alloys as a function of $x$ is shown in Fig.\ref{LocMagMom}. The
local magnetic moments found at the impurity sites in $MgC(Ni_{1-x}T_{x})_{3}$
alloys are inconsistent with the experiments. However, one may note
that the experimental characterizations on the effects of $3d$ transition-metal
impurities in disordered $MgCNi_{3}$ alloys have remained controversial. 

The appearance of local magnetic moments in disordered $MgC(Ni_{1-x}T_{x})_{3}$
alloys may be attributed to (i) an improper choice of the individual
atomic sphere radius, (ii) inadequate number of angular momentum states
in the basis set for the expansion of the wave functions, (iii) neglect
of local environment effects such as lattice relaxations and site
correlations, and (iv) an incorrect description of the exchange-correlation
potential. 

We have performed our calculations for disordered $MgC(Ni_{1-x}T_{x})_{3}$
alloys using different atomic sphere radii but still found local magnetic
moments at the impurity site, namely $Fe$ and $Co$. No local magnetic
moments were found at the $Cu$ site as well as $Ni$ site in $MgC(Ni_{1-x}T_{x})_{3}$
alloys. We have checked that the present calculations are well-converged
with respect to (a) angular momentum basis set, (b) number of energy
points in the contour for evaluating the Green's function and (c)
the number of $\mathbf{k}$-points in the irreducible part of the
Brillouin zone for $\mathbf{k}$-space integrations. 

The effects of lattice relaxation in disordered $MgC(Ni_{1-x}T_{x})_{3}$
alloys, if any, are expected to be very small. For example, the neglect
of the experimentally \cite{PRB-064509} determined deviation of around
$0.05$$\textrm{Å}$ of $Ni$ atoms from its ideal symmetry positions
in $MgCNi_{3}$ in earlier \cite{PRB-64-140507,PRB-100508,PRB-172507,PRL-027001}
as well as the present calculations show no magnetism in $MgCNi_{3}$.
Moreover, first- principles calculations for $3d$ impurities in a
$3d$ host suggest that the lattice relaxation effects would be less
important \cite{PRB-55-4157,PRB-39-930} in determining alloy energetics.
Usually, these effects are large if there is a large atomic size mismatch
between the host and the impurity atoms. 

The use of other exchange-correlation parametrization in our calculations
of $MgC(Ni_{1-x}T_{x})_{3}$ alloys still leads to the formation of
local magnetic moments at the $Fe$ and the $Co$ sites. Thus, the
possibility of inadequate description of the exchange-correlation
effects within the local-density approximation (LDA) cannot be ruled
out, especially when spin-fluctuation effects are important. For example,
$FeAl$, $Ni_{3}Ga$, $Sr_{3}Ru_{2}O_{7}$, $SrRhO_{3}$, and $NaCo_{2}O_{4}$
are paramagnets, but the LDA-based first-principles calculations show
a magnetic moment of $0.70$$\mu_{B}$ \cite{PRB-153106}, $0.79\,\mu_{B}$
\cite{PRL-92-147201}, $0.64\,\mu_{B}$ \cite{PRB-165101}, $0.83$$\mu_{B}$
\cite{PRB-54507}, and $0.35$$\mu_{B}$ \cite{PRB-13397,PRB-020503},
respectively. The Corrections to these overestimated magnetic moments
are, however, incorporated on a phenomenological level using the fixed-spin
moment method \cite{JPF-14-129}. 

We use a similar approach to study the incipient magnetism in the
disordered $MgC(Ni_{1-x}T_{x})_{3}$ alloys. Further, Singh and Mazin
\cite{PRB-64-140507} have anticipated spin-fluctuations in $MgCNi_{3}$
based on the Stoner model of itinerant magnetism. The overestimation
of magnetic moments in the ferromagnetic state of the alloys within
the LDA can be considered as an indicator of spin-fluctuations in
the system. However, for this to be an effective indicator, one needs
to rule out the competing states like anti-ferromagnetism. Empirically,
it has been suggested \cite{Int-J-QChem} that anti-ferromagnetism
would step in if the paramagnetic density of states has $E_{F}$ in
the non-bonding region. For those systems, where $E_{F}$ is in the
anti-bonding region, exchange splitting of bands leading to ferromagnetism
is more likely. In $MgCNi_{3}$, the $E_{F}$ is in the anti-bonding
region, and hence it is closer to ferromagnetic instability rather
than to an anti-ferromagnetic one. The absence of nesting features
in the Fermi surface also rules out anti-ferromagnetism in $MgCNi_{3}$\cite{PRB-100508,PRB-180510}.

\begin{figure}[h]
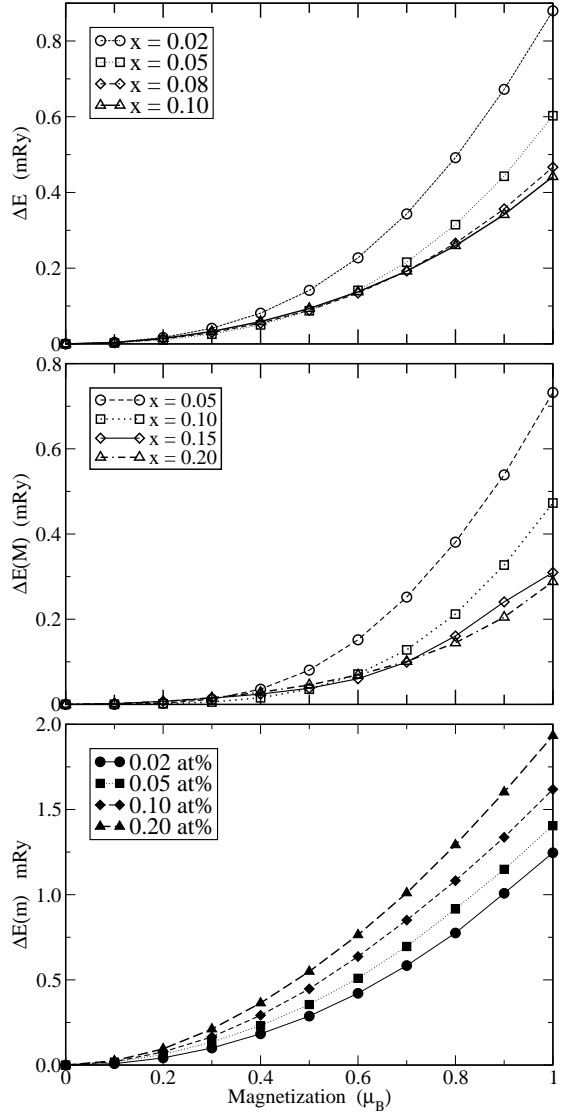

\includegraphics[%
  clip,
  scale=0.3]{fig2a.eps}

\includegraphics[%
  clip,
  scale=0.3]{fig2b.eps}

\includegraphics[%
  clip,
  scale=0.3]{fig2c.eps}

\caption{\label{FSM-DE-Fe}The magnetic energy $\Delta E(M)$ as a function
of magnetization $M$ for $MgC(Ni_{1-x}Fe_{x})_{3}$ (upper panel),
$MgC(Ni_{1-x}Co_{x})_{3}$(middle panel ), and $MgC(Ni_{1-x}Cu_{x})_{3}$(lower
panel) alloys, calculated at their respective equilibrium lattice
constants . The curves in the figure correspond to different concentrations
$x$ in $at$\% as indicated. }
\end{figure}

\begin{figure}[h]
\includegraphics[%
  clip,
  scale=0.3]{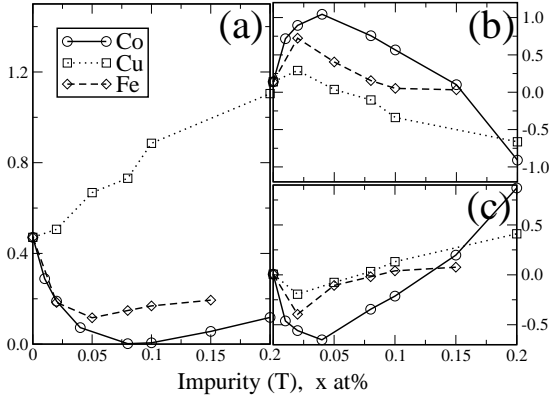}

\caption{\label{FSM-Coeff} The Ginzburg-Landau coefficients (a) $a_{2}$
in units of $10^{3}.\frac{T}{\mu_{B}}$, (b) $a_{4}$ in units of
$10^{3}.\frac{T}{\mu_{B}^{3}}$, and (c) $a_{6}$ in units of $10^{3}.\frac{T}{\mu_{B}^{5}}$,
as a function of $x$ in $MgC(Ni_{1-x}Fe_{x})_{3}$, $MgC(Ni_{1-x}Co_{x})_{3}$,
and $MgC(Ni_{1-x}T_{x})_{3}$ alloys, calculated at their respective
equilibrium lattice constants as described in the text. }
\end{figure}

We have calculated the magnetic energy $\Delta E(M)$ as a function
of $M$ for various values of $x$ for $MgC(Ni_{1-x}T_{x})_{3}$ alloys
at the equilibrium lattice constants using the fixed-spin moment method.
The calculated results of $\Delta E(M)$ as a function of $M$ for
$MgC(Ni_{1-x}Fe_{x})_{3}$, $MgC(Ni_{1-x}Co_{x})_{3}$ and $MgC(Ni_{1-x}Cu_{x})_{3}$
alloys are shown in Fig.\ref{FSM-DE-Fe}. It can be seen that the
curve $\Delta E(M)$ is rather flat near $M$=$0$ for small values
of $x$ in all the alloys and could be fit very well with the form
${\displaystyle {\displaystyle \sum_{n>0}\frac{1}{2n}a_{2n}M^{2n}}}$
for $n=4$. 

The changes in the Ginzburg-Landau coefficients $a_{2n}$ as a function
of $x$ are shown in Fig.\ref{FSM-Coeff}. From Fig.\ref{FSM-Coeff},
we find that $a_{2}$ decreases monotonically with increasing $x$
for $MgC(Ni_{1-x}Fe_{x})_{3}$ and $MgC(Ni_{1-x}Co_{x})_{3}$ alloys
in the range $0.0<x<0.2$. Since $a_{2}$'s are positive for the $MgC(Ni_{1-x}T_{x})_{3}$
alloys, one finds that the fixed-spin moment method in conjunction
with Ginzburg-Landau coefficients predicts the ground state of these
alloys to be paramagnetic, in agreement with the experiments. The
variation in $a_{4}$ and $a_{6}$ are also important as negative
values would indicate a metastable ground state. In the renormalization
approach to include spin-fluctuation effects from the first-principles,
the coefficients of the lower powers of $M$ are corrected by means
of higher-order coefficients. However, in the present case, one may
expect that the trend in the variation of $a_{2}$ would remain more
or less similar, as the variation in $a_{4}$ and $a_{6}$ with respect
to $x$ compliment each other. If one takes $a_{2}$=$0$ to be the
magnetic line of instability then as an increasing function of $x$
in $MgC(Ni_{1-x}T_{x})_{3}$ alloys, one finds that both $Fe$ and
$Co$ substitutions in low concentration enhance the incipient magnetic
traits when compared to $MgCNi_{3}$ itself. For $MgCNi_{3}$, the
Ginzburg-Landau coefficients $a_{2}$ and $a_{4}$ were determined
to be $470.5$ $\frac{T}{\mu_{B}}$ and $139.2$ $\frac{T}{\mu_{B}^{3}}$
, respectively.

\begin{figure}[h]
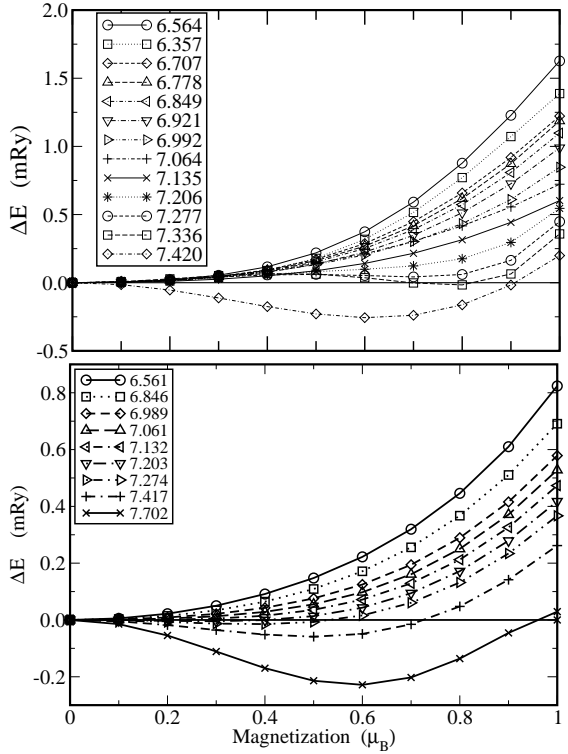

~

~

\includegraphics[%
  clip,
  scale=0.3]{fig4a.eps}

\includegraphics[%
  clip,
  scale=0.3]{fig4b.eps}

\caption{\label{Vol-Fe} The magnetic energy $\Delta E(M)$ as a function
of magnetization $M$ for $MgC(Ni_{0.95}Fe_{0.05})_{3}$ (upper panel)
and $MgC(Ni_{0.90}Co_{0.10})_{3}$(lower panel ) alloys, calculated
at various lattice constants as indicated. }
\end{figure}

Finally, we describe the effects of volume on the magnetic properties
of $MgC(Ni_{1-x}Fe_{x})_{3}$ and $MgC(Ni_{1-x}Co_{x})_{3}$ alloys.
To understand the magneto-volume relationship in $MgC(Ni_{1-x}Fe_{x})_{3}$
and $MgC(Ni_{1-x}Co_{x})_{3}$ alloys, we have carried out fixed-spin
moment calculations over a range of volume for $MgC(Ni_{0.95}Fe_{0.05})_{3}$
and $MgC(Ni_{0.90}Co_{0.10})_{3}$ alloys, respectively. We have chosen
these concentrations because the calculated local magnetic moments
are maximum at these concentrations. 

The calculated magnetic energy as a function of magnetization for
various lattice constants is shown in Fig. \ref{Vol-Fe}. Fig. \ref{Vol-Fe}
indicates the possibility of a ferromagnetic state for $MgC(Ni_{0.95}Fe_{0.05})_{3}$
and $MgC(Ni_{0.90}Co_{0.10})_{3}$ alloys. We also find that the paramagnetic
state becomes stable at smaller lattice constants for both $Fe$ and
$Co$ substitutions in $MgCNi_{3}$. However, for $Cu$ substitutions
(not shown), no such volume dependency on magnetism was observed.
As indicated earlier, following the rigid band picture, electron dopings
in the form of $Cu$ substitutions in $MgCNi_{3}$ decreases the density
of states at $E_{F}$ thus making it paramagnetic. 

In conclusion, the first-principles, local-density-functional based
calculations for substitutionally disordered $Fe$ and $Co$ impurities
in the $Ni$ sub-lattice of $MgCNi_{3}$, in low concentrations, show
that incipient magnetism resides in these materials. The overestimation
of the calculated magnetic properties points to the limitations of
the local-density approximation. Using a phenomenological approach
based on Ginzburg-Landau coefficients and the fixed-spin moment method,
we have shown that the disordered $MgC(Ni_{1-x}T_{x})_{3}$ ($T$$\equiv$$Fe$,
$Co$ or $Cu$) alloys remain paramagnetic. At expanded volumes, we
also find the possibility of a ferromagnetic state for $MgC(Ni_{0.95}Fe_{0.05})_{3}$
and $MgC(Ni_{0.90}Co_{0.10})_{3}$ alloys. 

One of us (PJTJ) would like to thank Dr. Andrei V. Ruban for discussions
on the theory and implementation of his KKR-ASA-CPA code. Discussions
with Dr. Igor I. Mazin on the various aspects of materials properties
and on some of his previous works are gratefully acknowledged.


\begin{thebibliography}{1}
\bibitem{PRB-64-140507}D. J. Singh and I. I. Mazin, Phys. Rev. B 64, 140507 (2001)
\bibitem{PRL-257601}P. M. Singer, T. Imai, T. He, M. A. Hayward, and R. J. Cava, \textbf{}Phys.
Rev. Lett. 87, 257601 (2001)
\bibitem{Nature-411-54}T. He, K.A. Regan, M.A. Hayward, A.P. Ramirez, Y. Wang, P. Khalifah,
T. He, J.S. Slusky, N. Rogado, K. Inumaru, M.K. Haas, H.W. Zandbergen,
N.P. Ong, and R.J. Cava, Nature, 411, 54 (2001)
\bibitem{PRL-027001}H. Rosner, R. Weht, M. D. Johannes, W. E. Pickett, and E. Tosatti,
Phys. Rev. Lett. 88, 027001 (2002)
\bibitem{PRB-172507}J. H. Kim, J. S. Ahn, J. Kim, Min-Seok Park, S. I. Lee, E. J. Choi,
and S.-J. Oh, Phys. Rev. B 66, 172507 (2002) 
\bibitem{PRB-100508}S. B. Dugdale and T. Jarlborg, \textbf{}Phys. Rev. B 64, 100508 (2001)
\bibitem{PRB-064510}T. G. Kumary, J. Janaki, A. Mani, S. M. Jaya, V. S. Sastry, Y. Hariharan,
T. S. Radhakrishnan, and M. C. Valsakumar, Phys. Rev. B 66, 064510
(2002)
\bibitem{SSC-132-379}T. Klimczuk and R.J. Cava, Solid State Commun 132, 379 (2004) 
\bibitem{SSC-491}M. A. Hayward, M. K. Haas, A. P. Ramirez, T. He, K. A. Regan, N. Rogado,
K. Inumaru and R. J. Cava, \textbf{}Solid State Commun. 119, 491 (2001) 
\bibitem{BJP-32-755}M. Alzamora, D. R. Sanchez, M. Cindra and E. M. Baggio-Saitovitch,
Brazilian J. Phys 32, 755 (2002)
\bibitem{PRB-064503}A. Das and R. K. Kremer, \textbf{}Phys. Rev. B 68, 064503 (2003)
\bibitem{SSC-122-269}C. M. Granada, C. M. da Silva and A. A. Gomes, \textbf{}Solid State
Communications 122, 269 (2002) 
\bibitem{PRB-064525}I. G. Kim, J. I. Lee, and A. J. Freeman, \textbf{}Phys. Rev. B 65,
064525 (2002)
\bibitem{PRB-66-024520}I. R. Shein, A. L. Ivanovskii, E. Z. Kurmaev, A. Moewes, S. Chiuzbian,
L. D. Finkelstein, M. Neumann, Z. A. Ren, and G. C. Che Phys. Rev.
B 66, 024520 (2002)
\bibitem{PhysicaC-154}X. Zheng, Y Xu, Zhi Zeng and E. Baggio-Saitovitch, Physica C 408-410,
154 (2004)
\bibitem{JAP-91-8504}J. L. Wang, Y. Xu , Z. Zeng, Q. Q. Zheng, H. Q. Lin, J. Appl. Phys
91, 8504 (2002)
\bibitem{ProgMater-27-1}J. S. Faulkner, Prog. Mater. Sci. 27, 1 (1982)
\bibitem{PRB-45-13244}J. P. Perdew and Y. Wang, Phys. Rev. B 45, 13244 (1992)
\bibitem{PRL-55-600}N. E. Christensen and S. Satpathy, Phys. Rev. Lett. 55, 600 (1985).
\bibitem{PRB-51-5773}P. A. Korzhavyi, A. V. Ruban, I. A. Abrikosov, and H. L. Skriver Phys.
Rev. B 51, 5773 (1995)
\bibitem{CMC-15-119}A. V. Ruban and H. L. Skriver Computational Materials Science, 15,
119 (1999)
\bibitem{PRB-66-024201}A. V. Ruban and H. L. Skriver Phys. Rev. B 66, 024201 (2002)
\bibitem{PRB-66-024202}A. V. Ruban, S. I. Simak, P. A. Korzhavyi, and H. L. Skriver Phys.
Rev. B 66, 024202 (2002)
\bibitem{PRB-56-9319}I. A. Abrikosov, S. I. Simak, B. Johansson, A. V. Ruban, and H. L.
Skriver Phys. Rev. B 56, 9319 (1997)
\bibitem{PRB-064509}A. Yu. Ignatov, L. M. Dieng, T. A. Tyson, T. He, and R. J. Cava, \textbf{}Phys.
Rev. B 67, 064509 (2003)
\bibitem{PRB-55-4157}N. Papanikolaou, R. Zeller, P. H. Dederichs and N. Stefanou, Phys.
Rev. B 55, 4157 (1997)
\bibitem{PRB-39-930}B. Drittler, M. Weinert, R. Zeller and P. H. Dederichs, Phys. Rev.
B 39, 930 (1989)
\bibitem{PRB-153106}I. I. Mazin, L. Chioncel and A. I. Lichtenstein, Phys. Rev. B 67,
153106 (2003). \textbf{}
\bibitem{PRL-92-147201}A. Aguayo, I. I. Mazin, and D. J. Singh\textbf{,}Phys. Rev. Lett.
92, 147201 (2004) 
\bibitem{PRB-165101}D. J. Singh and I. I. Mazin, \textbf{}Phys. Rev. B 63, 165101 (2001)
\bibitem{PRB-54507}D. J. Singh, \textbf{}Phys. Rev. B 67, 054507 (2003). 
\bibitem{PRB-13397}D.J. Singh, \textbf{}Phys. \textbf{}Rev. B 61, 13397 (2000)
\bibitem{PRB-020503}D.J. Singh, \textbf{}Phys. Rev. B 68, 020503 (2003)
\bibitem{JPF-14-129}K Schwarz and P Mohn, J. Phys. F: Met. Phys. 14, L129 (1984) 
\bibitem{Int-J-QChem}R. Dronskowski, Int. J. Quant. Chem. 96, 89 (2003)
\bibitem{PRB-180510}J. H. Shim, S. K. Kwon, and B. I. Min, \textbf{}Phys. Rev. B 64, 180510
(2001)\end{thebibliography}
\end{document}